\documentclass[letterpaper, 10 pt, conference]{ieeeconf}  

\IEEEoverridecommandlockouts                              

\usepackage{cite}
\usepackage{amsmath,amssymb,amsfonts}
\usepackage{algorithmic}
\usepackage{graphicx}
\usepackage{textcomp}
\usepackage{xcolor}
\usepackage{booktabs}
\usepackage{mathtools}
\usepackage{subcaption}
\usepackage{tabularx,booktabs,array,makecell, threeparttable}

\title{\LARGE \bf
Generative versus Discriminative Approaches for Class-Incremental Learning of EMG Signals: Effectiveness of Scale Mixture Modeling
}

\author{Seitaro Yoneda$^{1}$ Suguru Kanoga$^{2}$, and Akira Furui$^{1}$
\thanks{This work was partially supported by JSPS KAKENHI Grant Number JP23H0343800.}%
\thanks{$^{1}$Seitaro Yoneda and Akira Furui are with Graduate School of Advanced Science and Engineering, Hiroshima University, Higashi-hiroshima, Japan
        (e-mail: seitaroyoneda@hiroshima-u.ac.jp).}%
\thanks{$^{2}$Suguru Kanoga is with National Institute of Advanced Industrial Science and Technology (AIST), Koto-ku, Tokyo, Japan.}%
}

\begin{document}

\maketitle
\thispagestyle{empty}
\pagestyle{empty}

\begin{abstract}
In electromyogram (EMG)-based motion recognition, it is impractical to predefine all motions that may be required during deployment, necessitating class-incremental learning that sequentially adds new motion classes.
The primary challenges in class-incremental learning are catastrophic forgetting, where previously acquired knowledge is overwritten when learning new classes, and the memory cost of retaining past data to counteract it.
In particular, for EMG-based motion recognition intended for edge devices with limited computational resources, it is essential to suppress catastrophic forgetting and maintain low memory cost.
In this paper, we conducted a comparative evaluation of eight class-incremental learning methods spanning generative and discriminative approaches, including both deep and non-deep learning methods, for EMG signal classification.
Using four datasets, we evaluated each method in terms of classification accuracy, backward transfer, and memory cost.
The results demonstrated that deep learning-based methods suffered significant accuracy degradation from catastrophic forgetting as the number of tasks increased, whereas generative models maintained stable accuracy with low memory cost. 
Among generative models, the scale mixture classification model (SMCM), which captures EMG signal variability, achieved the most favorable accuracy–memory trade-off while effectively suppressing catastrophic forgetting across all datasets.
\end{abstract}

\section{Introduction}

Electromyogram (EMG) signals are bioelectrical signals generated during muscle contraction, measurable non-invasively from the skin surface via surface electrodes.
Estimating human motion intentions from EMG signals using machine learning has enabled various applications, including myoelectric prosthetic hands and interfaces for information devices~\cite{Fougner2012-wj}.

In EMG-based motion recognition, conventional machine learning models typically learn from data of all predefined motions in a batch manner.
However, in practical applications, it is impractical to predefine all motions that may be required in the future; systems must support the addition of new motion classes during deployment while preserving recognition of previously learned ones.
Such a framework is referred to as ``class-incremental learning''~\cite{van-de-Ven2022-uc}.

Class-incremental learning presents two primary challenges: catastrophic forgetting and the memory cost of countermeasures against it.
Catastrophic forgetting is a phenomenon in which previously acquired knowledge is overwritten when training on new class data alone, resulting in significant degradation in recognition accuracy for existing classes.
A common approach to suppress catastrophic forgetting is to retain past data and replay it during subsequent training; however, as the number of classes increases, the amount of retained data grows, leading to increased memory cost.
In particular, for EMG-based interfaces intended to operate on edge devices with limited computational resources, simultaneously suppressing catastrophic forgetting and maintaining low memory cost is essential.

Class-incremental learning methods can be broadly categorized into generative and discriminative models. 
Generative models estimate class-conditional probability distributions and classify data via Bayes' rule, with parameters estimated independently per class; thus, the addition of new classes does not require modification of existing class models. 
Among discriminative models, deep learning-based approaches have been extensively studied owing to their ability to extract effective features from data; however, they update the shared feature space when learning new classes, which can disrupt the representations of previously learned classes~\cite{LIU2025110943}. 
Various countermeasures, including replay, regularization, and architectural approaches~\cite{de2021continual}, have been proposed, but none eliminates this structural dependency on the shared feature space. 
This structural asymmetry raises the question of whether generative models with class-independent parameters offer systematic advantages over discriminative approaches for class-incremental learning under memory-constrained conditions.

Research on class-incremental learning for EMG signal classification remains limited~\cite{Campbell2025-qk}. 
Existing work has primarily employed replay-based deep learning methods~\cite{kanoga_embc,Lin2025-ue}. 
Although non-deep learning methods such as linear discriminant analysis (LDA)~\cite{Gu2018-pg} and support vector machine (SVM)~\cite{9871311} have been widely used in EMG signal classification and applied to continual learning for inter-session adaptation, they have not been systematically evaluated within the framework of class-incremental learning framework. 
More critically, no study has conducted a systematic cross-category comparison to examine whether the structural advantages described above hold in the EMG domain, nor has memory cost been evaluated assessed with a view to edge device deployment.

The authors have previously proposed a generative model called the scale mixture classification model (SMCM) for EMG pattern recognition~\cite{FURUI2021115644}. 
SMCM possesses three structural properties relevant to class-incremental learning: (i) each class is modeled by a completely independent set of parameters, so the addition of new classes does not affect existing class models; (ii) learned knowledge is compressed into posterior distributions of parameters, removing the need to store any past training data; and (iii) signal variability is captured through scale mixture distributions, which provide heavier-tailed representations than the Gaussian distributions assumed in methods such as streaming linear discriminant analysis (SLDA)~\cite{Hayes_2020_CVPR_Workshops}. 
These properties make SMCM a strong candidate for class-incremental learning
under memory-constrained conditions; however, their effectiveness has not been empirically validated in this context.

This study therefore conducts a systematic cross-category comparison to examine two questions: whether generative models with class-independent parameters exhibit systematic advantages over discriminative approaches in both forgetting suppression and memory efficiency; and whether SMCM's scale mixture distributions yield additional accuracy improvements over Gaussian-based generative models.
The contributions of this study are as follows:
\begin{itemize}
  \item A systematic comparison of eight class-incremental learning methods spanning generative and discriminative approaches for EMG signal classification, providing empirical verification of the hypothesized structural advantages of generative models
  \item Demonstration that generative models consistently outperform deep learning-based approaches across four diverse datasets, revealing the limitations of shared feature-space updates for EMG-based class-incremental learning
  \item Practical design insights for class-incremental EMG interfaces on edge device, derived from a multifaceted evaluation of accuracy, forgetting, and memory cost
\end{itemize}

\section{Methods}

\subsection{Problem Setting}

\begin{figure}
    \centering
    \includegraphics[width=\linewidth]{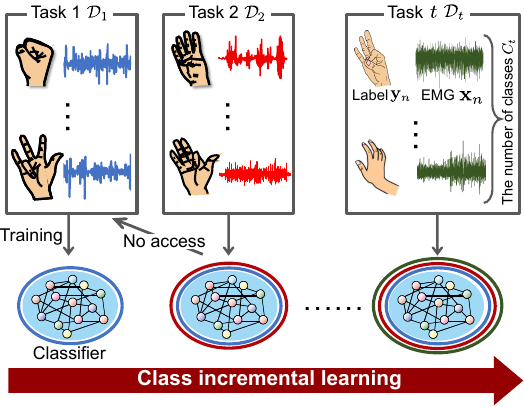}
    \caption{Schematic overview of class-incremental learning for EMG classification. The classifier learns sequentially from tasks $\mathcal{D}_1$, $\mathcal{D}_2$, $\ldots$, $\mathcal{D}_t$, each containing new motion classes, without access to past task data.}
    \label{fig:overview}
\end{figure}
Fig.~\ref{fig:overview} illustrates the class-incremental learning framework for EMG classification.
Let $\mathbf{x}_n \in \mathbb{R}^{D}$ denote the $D$-dimensional EMG feature vector and $\mathbf{y}_n \in \{0,1\}^{C}$ denote the corresponding one-hot class label, where $C$ is the number of classes.
Given a dataset $\mathcal{D} = \{(\mathbf{x}_n, \mathbf{y}_n)\}_{n=1}^N$ with $C$ classes, we partition $\mathcal{D}$ into $T$ tasks, where each task $\mathcal{D}_t$ ($t = 1, \ldots, T$) consists of data from $C_t$ new classes. 
The classifier learns sequentially from task $t = 1$; during the learning of task $t$, only the current task data $\mathcal{D}_t$ is available, without access to past task data $\mathcal{D}_1, \ldots, \mathcal{D}_{t-1}$. 
After learning task $t$, the classifier is required to perform classification over all $\sum_{i=1}^{t} C_i$ classes learned up to that point.

\subsection{Scale Mixture Classification Model (SMCM)}
SMCM is a generative model that estimates the class-conditional distribution 
$p(\mathbf{x}_n | \mathbf{y}_n)$ and infers the class posterior probability $p(\mathbf{y}_n | \mathbf{x}_n)$ via Bayes' rule. 
By weighting the variance with a latent variable $u_{n,c} \in \mathbb{R}^{+}$ following an inverse-gamma distribution, SMCM captures EMG signal variability through a scale mixture 
distribution~\cite{FURUI2021115644}. 
The class-conditional model for class $c$ is defined as follows:
\begin{align}
  p(&\mathbf{x}_n| y_{n,c} = 1)\nonumber\\ &= \int \mathcal{N}(\mathbf{x}_n | \boldsymbol{\mu}_c, u_{n,c} \boldsymbol{\Sigma}_c) \times \mathrm{IG}\left(u_{n,c} \middle| \frac{\nu_c}{2}, \frac{\nu_c}{2}\right) du_{n,c}, \label{eq:smcm_likelihood} \\
  p(&\mathbf{y}_n) = \mathrm{Cat}(\mathbf{y}_n | \boldsymbol{\pi}) = \prod_{c=1}^{C} \pi_c^{y_{n,c}}, \label{eq:smcm_prior}
\end{align}
where $\boldsymbol{\mu}_c \in \mathbb{R}^{D}$ and $\boldsymbol{\Sigma}_c \in \mathbb{R}^{D \times D}$ represent the mean vector and covariance matrix of class $c$, respectively.
The parameter $\nu_c \in \mathbb{R}^{+}$ controls the degree of uncertainty in the variance.
In addition, $\boldsymbol{\pi} = \{\pi_c\}$ denotes the mixing coefficients satisfying $\sum_{c=1}^{C} \pi_c = 1$.

SMCM places conjugate prior distributions on the model parameters $\boldsymbol{\Theta} = \{\boldsymbol{\mu}_c, \boldsymbol{\Sigma}_c, \boldsymbol{\pi}\}$. 
After observing the data $\mathcal{D}_t$ at task $t$, the posterior distribution $p(\boldsymbol{\Theta}, \mathbf{U}|\mathcal{D}_{t})$ over the model parameters and the latent variables $\mathbf{U}=\{u_{nc}\}$ is approximated via variational inference based on the mean-field assumption. 
The detailed definitions of the prior distributions and the derivation of the update rules can be found in~\cite{FURUI2021115644}. 
Owing to the conjugacy of the priors, the variational posterior distributions take the following forms:
\begin{align}
  q(\boldsymbol{\mu}_c, \boldsymbol{\Sigma}_c | \mathcal{D}_t) &= \mathcal{N}(\boldsymbol{\mu}_c | \mathbf{m}_c, \beta_c^{-1} \boldsymbol{\Sigma}_c) \, \mathrm{IW}(\boldsymbol{\Sigma}_c | \mathbf{W}_c, \eta_c), \label{eq:smcm_post_musigma} \\
  q(\boldsymbol{\pi} | \mathcal{D}_t) &= \mathrm{Dir}(\boldsymbol{\pi} | \boldsymbol{\alpha}), \label{eq:smcm_post_pi}
\end{align}
where $\mathbf{m}_c$, $\beta_c$, $\mathbf{W}_c$, $\eta_c$, and $\boldsymbol{\alpha}$ are the hyperparameters of the variational posterior distributions, updated from the observed data.

Because the hyperparameters $(\mathbf{m}_c, \beta_c, \mathbf{W}_c, \eta_c)$ of each class $c$ depend only on the data of that class, adding new classes $c^{\prime}$ at task $t+1$ requires estimating only $(\mathbf{m}_{c^\prime}, \beta_{c^\prime}, \mathbf{W}_{c^\prime}, \eta_{c^\prime})$ from $\mathcal{D}_{t+1}$; the hyperparameters of existing classes remain unchanged.
Since all learned knowledge is compressed into these hyperparameters, no past data need to be retained.

At inference, the predictive distribution for a new observation $\mathbf{x}_n$ is computed from the expectations under the variational posterior distributions of each learned class $c$:
\begin{align}
  p(\mathbf{x}_n | y_{n,c} = 1, \mathcal{D}_1, \ldots, \mathcal{D}_t) \approx p(\mathbf{x}_n | \langle \boldsymbol{\mu}_c \rangle, \langle \boldsymbol{\Sigma}_c \rangle, \nu_c), \label{eq:smcm_pred}
\end{align}
where $\langle \boldsymbol{\mu}_c \rangle = \mathbf{m}_c$ and $\langle \boldsymbol{\Sigma}_c \rangle = (\eta_c - D - 1)^{-1} \mathbf{W}_c$.

The predicted label $\hat{c}_n$ is assigned to the class with the highest posterior probability obtained via Bayes' rule:
\begin{align}
  \hat{c}_n = \underset{c}{\arg\max}\ p(y_{n,c} = 1 | \mathbf{x}_n, \mathcal{D}_1, \ldots, \mathcal{D}_t). \label{eq:smcm_decision}
\end{align}

\subsection{Compared Methods}

\subsubsection{Generative Models}

SLDA~\cite{Hayes_2020_CVPR_Workshops} is a probabilistic model that represents class-conditional distributions as Gaussian distributions.
The class-conditional model for class $c$ is defined as:
\begin{align}
  p(\mathbf{x}_n | y_{n,c} = 1) = \mathcal{N}(\mathbf{x}_n | \boldsymbol{\mu}_c, \boldsymbol{\Sigma}), \label{eq:slda}
\end{align}
where $\boldsymbol{\Sigma} \in \mathbb{R}^{D \times D}$ is a covariance matrix shared among all classes.
Let $\boldsymbol{\Sigma}_{\mathrm{old}}$ denote the covariance matrix estimated using $N$ data points up to task $t-1$, and suppose that task $t$ provides $\mathcal{D}_t$ containing $N_t$ new data points.
The updated covariance matrix $\boldsymbol{\Sigma}_{\mathrm{new}}$ is then computed as the following weighted average:
\begin{align}
  \boldsymbol{\Sigma}_{\mathrm{new}} = \frac{N \boldsymbol{\Sigma}_{\mathrm{old}} + N_t \boldsymbol{\Sigma}_t}{N + N_t}, \label{eq:slda_update}
\end{align}
where $\boldsymbol{\Sigma}_t$ is the covariance matrix computed from the data of task $t$.

\subsubsection{Discriminative Models}


As a non-deep-learning discriminative method, we employed incremental support vector machine (ISVM)~\cite{NIPS2000_155fa095}, which extends SVM to the class-incremental learning framework to SVM.
ISVM retains support vectors after each task and combines them with new data during subsequent training to suppress catastrophic forgetting.

Deep learning-based class-incremental learning methods can be broadly categorized into three approaches: regularization, architectural, and replay. 
We selected representative methods from each category. 
Learning without forgetting (LwF)~\cite{8107520} is a regularization method that suppresses forgetting by constraining the output distributions or parameters of the previously trained model. 
Copy weights with re-init+ (CWR+)~\cite{MALTONI201956} is an architectural method that mitigates prediction bias toward new classes by adjusting the output layer weights.
For replay, we employed experience replay (ER) and average gradient episodic 
memory (AGEM)~\cite{NEURIPS2019_fa7cdfad,chaudhry2018efficient}, which retain a subset of past real data, and deep generative replay (DGR)~\cite{NIPS2017_0efbe980}, which generates pseudo-data using a variational autoencoder.

As baselines indicating the upper and lower bounds of accuracy for the deep learning methods, we defined JOINT and NONE~\cite{kanoga_embc}. 
JOINT trains the model using all data from previously encountered tasks, representing the upper bound. 
NONE fine-tunes the previously trained model using only the new task data, representing the lower bound without any forgetting countermeasures.

\begin{table}[t]
\centering
\caption{Summary of datasets used in the experiments}
\label{tab:datasets}
\scriptsize 
\setlength{\tabcolsep}{2.0pt} 
\renewcommand{\arraystretch}{1.05}
\begin{tabularx}{\columnwidth}{@{}l *{4}{>{\centering\arraybackslash}X}@{}}
\toprule
Dataset & \# Participants~($S$) & \# Motions~($C$) & \# Electrodes~($D$) & \# Trials~($T$) \\
\midrule
I   & 6  & 6  & 4  & 7 \\
II  & 8  & 14 & 8  & 4 \\
III & 40 & 10 & 4  & 5 \\
IV   & 10 & 16 & 12 & 6 \\
\bottomrule
\end{tabularx}
\end{table}

\subsection{Datasets}
We used four datasets: a private dataset~I of upper-limb motions~\cite{10394290} and three public datasets~II, III and~IV~\cite{Khushaba2013-va,Ozdemir2022-jo,Pizzolato2017-am}.
Table~\ref{tab:datasets} summarizes each dataset.
For dataset~IV, the last class was excluded from the original 17 classes to ensure an even number of classes for equal task splitting, resulting in 16 classes.

\begin{figure*}[t]
    \centering
    \includegraphics[width=\linewidth]{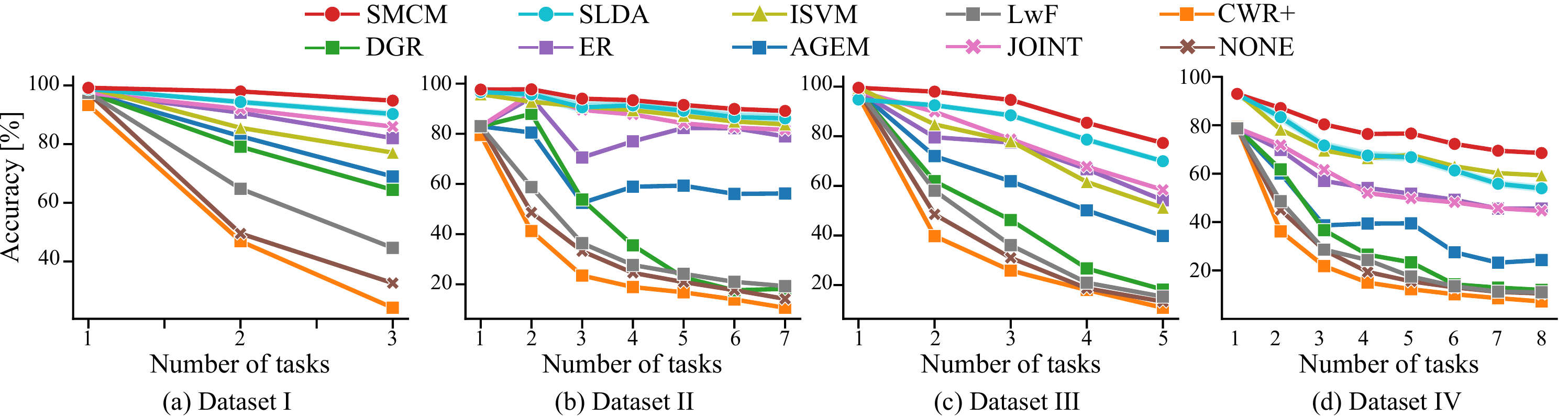}
    \caption{Average classification accuracy across tasks for all compared methods on each dataset. (a) Dataset~I. (b) Dataset~II. (c) Dataset~III. (d) Dataset~IV.}
    \label{fig:acc_task}
\end{figure*}

\subsection{Experimental Setup}

\subsubsection{Preprocessing}

To ensure that each method category operates under its established conditions, we applied preprocessing protocols from the respective literature. 
For the non-deep-learning methods (SMCM, SLDA, ISVM), feature extraction was performed by applying full-wave rectification and smoothing using a second-order Butterworth low-pass filter with a cutoff frequency of 2.0~Hz, following the standard pipeline for generative EMG classifiers~\cite{Yoneda2025-bg,FURUI2021115644,10394290}. 
For the deep learning methods, a window of 250~ms with 80\% overlap was applied, and the data were resized to $64 \times 64$ and treated as grayscale images, following~\cite{kanoga_embc}.

\subsubsection{Evaluation Protocol}

The performance was evaluated using leave-one-trial-out cross-validation on each dataset: one trial was used as test data and the remaining trials as training data, and this procedure was repeated for all trials.
The training data were partitioned into tasks with $C_t = 2$ classes per task, 
yielding $T = C/2$ tasks. 
For example, a dataset with $C = 6$ classes yields $T = 3$ tasks (task~1: classes 1--2, task~2: classes 3--4, task~3: classes 5--6).
Note that the class ordering was fixed throughout all experiments, although class ordering is known to affect the performance of class-incremental learning methods.

Each method was evaluated from three perspectives: classification accuracy, the degree of catastrophic forgetting, and memory cost.
For classification accuracy, we computed the average accuracy over all classes learned up to that point after the completion of each task $t$.

To quantify catastrophic forgetting, we employed backward transfer (BWT)~\cite{NIPS2017_f8752278}, which measures the impact of learning new tasks on the accuracy of previously learned tasks:
\begin{align}
  \mathrm{BWT} = \frac{1}{T-1} \sum_{i=1}^{T-1} (a_{T,i} - a_{i,i}), \label{eq:bwt}
\end{align}
where $a_{t,i}$ represents the classification accuracy on the test data of task $i$ after the completion of task $t$.
A BWT value of zero or above indicates that past knowledge is retained, whereas a large negative value indicates severe catastrophic forgetting.

Memory cost was defined as thetotal storage required by each method after the completion of all tasks.
Specifically, we summed the size of model parameters and, for methods that retain past data (e.g., ER, ISVM, JOINT), the buffer size, and the total was reported the total in bytes.

\begin{figure*}[t]
    \centering
    \includegraphics[width=1.0\linewidth]{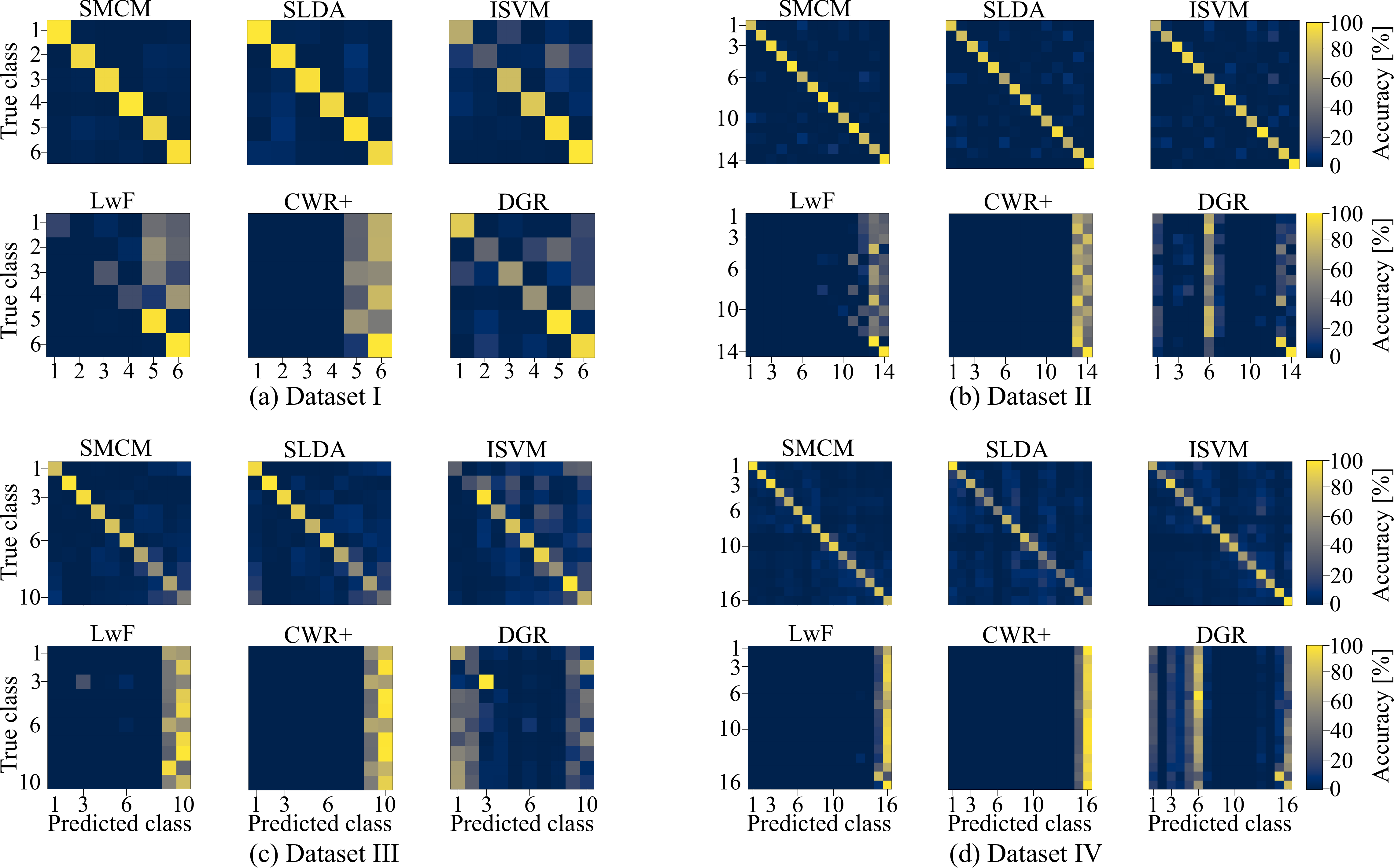}
    \caption{Confusion matrices after the final task for six representative methods on each dataset. (a) Dataset~I. (b) Dataset~II. (c) Dataset~III. (d) Dataset~IV.}
    \label{fig:confusion_matrix}
\end{figure*}

\subsubsection{Implementation Details}

For SMCM, the hyperparameters of the prior distributions were set to form weakly informative priors following~\cite{Yoneda2025-bg}.
The degree-of-freedom parameter, which controls the tail heaviness of the scale mixture distribution, was set to $\nu_0 = 1$.

ISVM was implemented using the SVC class with a radial basis function (RBF) kernel from scikit-learn~\cite{JMLR:v12:pedregosa11a}. 
The regularization parameter was set to 1, and the kernel bandwidth parameter $\gamma$ was set to $1/(D \cdot \mathrm{Var}(\mathbf{X}))$, where $\mathrm{Var}(\mathbf{X})$ denotes the variance of the training data.
Only the support vectors identified during training were retained for use in 
subsequent tasks.

For the deep learning methods, the model architecture was unified for fair comparison.
The feature extractor consisted of three convolutional layers with 8, 16, and 32 feature maps, each followed by batch normalization and ReLU activation.
The classifier comprised two fully connected layers with a ReLU activation function applied to the intermediate layer.
The Adam optimization ($\beta_1 = 0.9$, $\beta_2 = 0.999$) was used with a learning rate of $10^{-3}$, a mini-batch size of 64, and training for 400 epochs~\cite{kanoga_embc}.
For ER, a random subset of the training data equal to the mini-batch size was retained per class after each task for replay. 


\begin{figure}[t]
    \centering
    \includegraphics[width=1.0\linewidth]{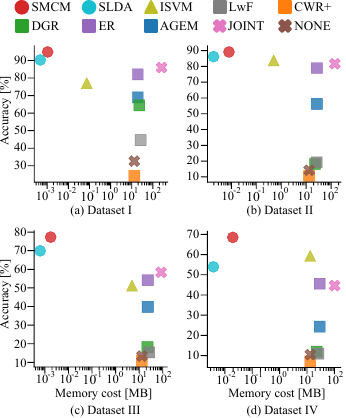}
    \caption{Average classification accuracy versus memory cost (log scale) for all compared methods on each dataset. (a) Dataset~I. (b) Dataset~II. (c) Dataset~III. (d) Dataset~IV.}
    \label{fig:acc_memory_cost}
\end{figure}

\begin{table*}[t]
\centering
\caption{Backward transfer (BWT) for each method on each dataset ($\uparrow$ higher is better).}
\label{tab:bwt}
\begin{threeparttable}
\begin{tabular}{l cc cccccc cc}
\toprule
& \multicolumn{2}{c}{Generative} & \multicolumn{6}{c}{Discriminative} & \multicolumn{2}{c}{Baseline} \\
\cmidrule(lr){2-3} \cmidrule(lr){4-9} \cmidrule(lr){10-11}
Dataset & SMCM & SLDA & ISVM & LwF & CWR+ & DGR & ER & AGEM & JOINT & NONE \\
\midrule
I   & \textbf{$-$0.033} & \underline{$-$0.048} & $-$0.322 & $-$0.796 & $-$0.739 & $-$0.424 & $-$0.090 & $-$0.414 & $-$0.052 & $-$0.981 \\
II  & \underline{$-$0.052} & $-$0.064 & $-$0.108 & $-$0.896 & $-$0.462 & $-$0.617 & \textbf{$-$0.049} & $-$0.463 & $0.061$  & $-$0.959 \\
III & \underline{$-$0.082} & \textbf{$-$0.078} & $-$0.409 & $-$0.937 & $-$0.551 & $-$0.532 & $-$0.125 & $-$0.536 & $-$0.061 & $-$0.962 \\
IV  & \underline{$-$0.079} & $-$0.119 & $-$0.206 & $-$0.793 & $-$0.300 & $-$0.368 & \textbf{$-$0.031} & $-$0.639 & $0.060$  & $-$0.791 \\
\bottomrule
\end{tabular}
\begin{tablenotes}
\item The best and second-best results among non-baseline methods are highlighted in \textbf{bold} and \underline{underline}, respectively.
\end{tablenotes}
\end{threeparttable}
\end{table*}

\section{Results}

\subsection{Classification Accuracy across Tasks}

Fig.~\ref{fig:acc_task} shows the average classification accuracy across tasks for all methods on each dataset.
SMCM maintained the most stable accuracy as the number of tasks increased across all datasets, achieving average classification accuracies at the final task of 94.84\% on Dataset~I, 89.12\% on Dataset~II, 77.29\% on Dataset~III, and 68.51\% on Dataset~IV.
SLDA also showed a small accuracy degradation as tasks increased; however, it was consistently below SMCM across all datasets, with the difference being particularly pronounced on Dataset~IV (SLDA: 53.94\%).
The deep learning-based methods, LwF, CWR+, DGR, and AGEM, exhibited substantial accuracy degradation with increasing tasks, despite incorporating measures against catastrophic forgetting.
By contrast, ER was the only deep learning-based method that suppressed accuracy degradation, owing to its retention and replay of a portion of past data.
ISVM exhibited less accuracy degradation compared to the deep learning methods; however, the difference from SMCM varied across datasets.

Fig.~\ref{fig:confusion_matrix} shows the confusion matrices of six representative methods (SMCM, SLDA, ISVM, LwF, CWR+, DGR) after the final task.
The color of each cell represents the proportion of the model's predicted label (horizontal axis) for a given true class label (vertical axis), with lighter colors indicating values closer to 1.0.
SMCM, SLDA and ISVM showed high diagonal values across all classes, confirming that the classification performance of past classes was maintained. 
By contrast, LwF, CWR+, and DGR showed low diagonal values for earlier classes with predictions biased toward later classes. 

\subsection{Backward Transfer}

Table~\ref{tab:bwt} presents the BWT for all methods on each dataset.
A larger negative BWT value indicates more severe catastrophic forgetting.
The generative models, SMCM and SLDA, exhibited BWT values ranging from $-0.119$ to $-0.033$ across all datasets, indicating that the knowledge of past classes was stably retained.
Among the deep learning methods, ER exhibited BWT values ranging from $-0.125$ to $-0.031$, comparable to those of SMCM.
By contrast, LwF showed BWT values of $-0.937$ to $-0.793$ across all datasets, demonstrating severe forgetting comparable to NONE, and the other methods also showed large negative values.
For the baselines, JOINT achieved positive BWT on some datasets, whereas NONE exhibited extremely large negative values ranging from $-0.981$ to $-0.791$ across all datasets.

\subsection{Accuracy--Memory Cost Trade-off}

Fig.~\ref{fig:acc_memory_cost} shows the relationship between memory cost and average classification accuracy for each method on each dataset.
SMCM and SLDA achieved high accuracy with low memory cost, positioned in the upper-left region of the plots.
SMCM achieved higher accuracy than SLDA with comparable memory cost.
Among the deep learning methods, LwF, CWR+, DGR, and AGEM exhibited substantially lower accuracy despite having higher memory cost than the generative models.
ER achieved accuracy comparable to SMCM; however, it required significantly higher memory cost owing to the retention of past data. 
JOINT demonstrated the highest accuracy but also the largest memory cost.

\section{Discussion}


Generative models that estimate class-conditional probability distributions exhibited a systematic advantage over discriminative models in suppressing catastrophic forgetting (Figs.~\ref{fig:acc_task},~\ref{fig:confusion_matrix} and Table~\ref{tab:bwt}).
In generative models, the decision boundary is determined via Bayes' rule from these distributions; thus, compared to discriminative models that re-learn the entire decision boundary, the impact on existing classes when adding new classes is limited. 
This structural property enables the avoidance of catastrophic forgetting without retaining or regenerating past data. 
By contrast, the deep learning-based methods exhibited accuracy degradation comparable to NONE, despite incorporating forgetting countermeasures. 
This is because deep learning models update the entire feature space when learning new classes, which is likely to disrupt the representations of previously learned classes~\cite{LIU2025110943}. 
The BWT values in Table~\ref{tab:bwt} corroborate this finding. 
Although ER exhibited favorable BWT, this resulted from the direct retention of past data for retraining rather than from a structural resistance to forgetting. 
ISVM maintained relatively stable performance, as it learns decision boundaries in the input space and is thus less susceptible to forgetting caused by feature space updates. 
These results highlight the limitations of deep learning-based forgetting countermeasures that involve feature space updates, suggesting that for class-incremental learning of EMG signals, designing generative models tailored to the signal characteristics is more effective than applying generic deep learning-based approaches.


Turning to the comparison within generative models, the accuracy gap between SMCM and SLDA can be attributed to two structural differences (Fig.~\ref{fig:acc_memory_cost}). 
First, SLDA employs a covariance matrix shared among all classes; thus, the update of the covariance matrix upon the addition of new classes can affect the decision boundaries of existing classes. 
By contrast, SMCM maintains completely independent parameters for each class, and the addition of new classes does not affect existing class models. 
Second, SMCM models EMG signal variability as uncertainty in variance, yielding a heavier-tailed distribution than the Gaussian assumed in SLDA. 
Under conditions with a large number of classes, such as Dataset~IV, the repeated updates of the shared covariance matrix in SLDA cause the impact on existing classes to accumulate, and the increased density of the feature space makes the appropriateness of the distributional assumption more influential on inter-class discrimination. 
The particularly pronounced accuracy difference on this dataset suggests that these two factors acted in combination. 
Although SLDA slightly outperformed SMCM in terms of BWT on Dataset~III, the forgetting suppression performance of SMCM was comparable to or better than that of SLDA across all datasets. 
Overall, the accuracy improvement afforded by scale mixture distributions was consistently confirmed across all datasets, with comparable or superior forgetting suppression, demonstrating that SMCM's scale mixture distributions provide additional advantages over Gaussian-based generative models.

Beyond accuracy and forgetting suppression, the trade-off with memory cost is critical for edge deployment.
Fig.~\ref{fig:acc_memory_cost} reveals a clear separation between methods that compress knowledge into parameters and those that retain past data. 
Generative models compress learned knowledge into model parameters rather than retaining past data: SMCM stores knowledge as posterior distributions of parameters, whereas SLDA retains sufficient statistics. 
This compression underlies the low memory cost of both models.
ER and ISVM also achieved accuracy comparable to that of SMCM; however, they incur substantially higher memory cost. 
ER, being a deep learning-based method, has a large number of model parameters and additionally requires retaining past data in a buffer, resulting in high memory cost. 
ISVM itself has a small model footprint; however, it must retain support vectors, so memory cost grows as the number of classes increases. 
Where sufficient memory is available, ER and ISVM remain viable options given their high classification accuracy.
Under the resource constraints typical of edge devices, however, SMCM offers the most favorable accuracy--memory trade-off among the methods evaluated.

\section{Conclusion}
This study conducted a systematic comparison of generative and discriminative approaches for class-incremental learning of EMG signals, evaluating eight methods across four datasets.  
The results using four datasets demonstrated that generative models with class-independent parameters consistently outperformed deep learning-based approaches in both forgetting suppression and memory efficiency.  
Among generative models, SMCM yielded additional accuracy improvements over the Gaussian-based SLDA across all datasets.
These findings suggest that generative models compressing knowledge into parameter distributions offer an effective design strategy for class-incremental EMG interfaces on resource-constrained edge devices.
In particular, SMCM achieved the most favorable accuracy–memory trade-off among the methods evaluated.

This study has several limitations. 
The experiments were limited to offline evaluation; thus, inference time and power consumption on edge devices remain unexamined. 
In addition, only a fixed two-class increment per task with a fixed class ordering was considered. 
Future work should include online evaluation on edge devices and environments with varying task sizes and randomized class orderings.

\bibliographystyle{IEEEtran} 
\bibliography{reference} 

\end{document}